\documentstyle{article}

         \def\ba{\begin{array}}
         \def\ea{\end{array}}
         \def\be{\begin{equation}}
         \def\bea{\begin{eqnarray}}
         \def\eea{\end{eqnarray}}
         \def\ee{\end{equation}}
         \def\L{{\Lambda }}
         \def\l{{\lambda}}

         \def\d{\partial}
         \def\v{\delta}
         \def\D{\Delta}
         \def\e{\epsilon}
         \def\a{\alpha}
         \def\b{\beta}
         \def\o{\omega}
         
         \def\bq{\begin{eqnarray}}
         \def\eq{\end{eqnarray}}
         \def\f{\!\!\!\!\!\!\!}
\begin{document}
\hfill
\vbox{
    \halign{#\hfil          \cr
            hep-th/9610169  \cr
            IPM-96-169      \cr
           } 
      }  
\vskip 10 mm
\leftline{ \Large \bf
           Derivation of theories: structures of the derived system} 
\vskip 7mm
\leftline{ \Large \bf
           in terms of those of the original system in classical mechanics} 

\vskip 15 mm
\leftline{ \bf Mohammad Khorrami$^{1,2,3,*}$, Amir Aghamohammadi$^{2,4}$}
\vskip 10 mm
{\it
\leftline{ $^1$ Department of Physics, Tehran University,
             North-Kargar Ave. Tehran, Iran. }
\leftline{ $^2$ Institute for Studies in Theoretical Physics and Mathematics,
           P.O.Box  5531, Tehran 19395, Iran. }
\leftline{ $^3$ Institute for Advanced Studies in Basic Sciences,
             P.O.Box 159, Gava Zang, Zanjan 45195, Iran. }
\leftline{ $^4$ Department of Physics, Alzahra University,
             Tehran 19834, Iran }
\leftline{ $^*$ E-Mail: mamwad@netware2.ipm.ac.ir}
  }
\leftline{PACS codes: 03.20.+i, 03.50.-z} 
\leftline{keywords: Derivation, Integrability, Symmetry, 
Classical field thoery }
\begin{abstract}
We present the technique of derivation of a theory to obtain an 
$(n+1)f$-degrees-of-freedom theory from an $f$-degrees-of-freedom theory and 
show that one can calculate all of the quantities of the derived theory from 
those of the original one. Specifically, we show that one can use this 
technique to construct, from an integrable system, other integrable
systems with more degrees of freedom.
\end{abstract}

\vskip 10 mm
\section{\bf Introduction} There are not many systems of more than one 
degrees of freedom, which can be manipulated easily. In this article, we 
propose a method to construct from a system, another one with more degrees 
of freedom. In this construction, almost anything which can be said about 
the original system has an analogue in the resulting system.

The main idea of this construction is based on the formal derivation of the 
entities of the original system with respect to a parameter, which may or 
may not explicitly appear in the original theory. One meaning of this is 
that the initial conditions of the system are functions of this parameter 
($\l$) \cite{Lax} and the solution of the original system depends on $\l$ 
through these initial conditions, and through the evolution of the system, 
which may depend on $\l$, if $\l$ appears explicitly in the original theory. 
Another way of viewing this is through the concept of contraction: consider 
two systems with parameters $\l$ and $\l +\D$. These two systems are 
independent of each other. One can write an action as the difference of the 
actions of the two system divided by $\D$ to discribe both systems. One can 
use one of these degrees of freedom and the difference of them divided by 
$\D$ a new set of varibales. This system, however, is equivalent to two 
copies of the original system. But if one lets $\D$ tend to zero, a 
well-defined theory of double number of variables is obtained, which can be 
nolonger decomposed to two independent parts. One can, however, solve this 
theory in terms of the solution of the original theory. This procedure is 
nothing but a contraction.

In section 2, we introduce the concept of derivation of a theory, and obtain 
the action, the equation of motion, and the solution of an $n$ times derived 
theory in terms of those of the original theory. In section 3, we do the 
same thing for the phase space of the derived theory, and obtain its momenta 
and Hamiltonian in terms of those of the original one. In section 4, it is 
shown that any symmetry, and any constant of motion of the original theory, 
results in a symmetry and a constant of motion of the derived one. In fact, 
any constant of motion of the original theory leads to $n+1$ constants of 
motion of its $n$ times derived theory. This fact has an important 
consequence, which we show in section 5: any theory derived from an 
integrable theory is integrable. There are several techniques of obtaining 
and studying integrable systems \cite{Lax,Tod}. The technique of derivation 
enables one to obtain an integrable system of $(n+1)f$ degrees of freedom 
from an integrable system of $f$ degrees of freedom. Specifically, as any 
system of one degrees of freedom is integrable, one can easily construct 
integrable systems of $n+1$ degrees of freedom. At last, it is easy to see 
that this technique is applicable to classical field theories as well. A 
group of $1+1$ dimensional integrable field theories are those which have a 
Lax structure \cite{Zak}. In section 7, we show that the derivation of such 
a theory has a Lax structure as well, and hence is integrable. We also 
obtain its Lax pair in terms of that of the original theory. 

This technique is applicable to quantum systems as well. There, other 
interesting questions arise. We will address this problem in a future work 
\cite{Agh}.

\section{\bf Lagrangian formulation, equation of motion, 
and its solution} 
Consider the Lagrangian 
\be \label{1*}
L^{(0)}= L(x, \dot x, \l)
\ee
where $x$ denotes the coordinte(s) of the configuration space and $\l$ 
is some parameter. Now differentiate this Lagrangian with respect to $\l$,
treating $x$ as a function of $\l$. In this way, one finds another 
Lagrangian
\be
L^{(1)} ={\d L^{(0)}\over \d \l}+ {\d L^{(0)}\over \d x}{\d x\over \d \l}
+{\d L^0\over \d \dot x}{\d \dot x\over \d \l}.
\ee
Fixing $\l$ and introducing 
\be 
x^{(0)} =x,\qquad x^{(1)}={\d x\over \d \l},
\ee
one can see that this new Lagrangian is in fact a function of two set 
of variables
\be
L^{(1)}(x^{(0)},x^{(1)},  \dot x^{(0)}, \dot x^{(1)}, \l)=                 
{\d L^{(0)}(x, \dot x, \l)\over \d \l}+ {\d L^{(0)}(x, \dot x, \l)\over \d x}
x^{(1)}+{\d L^{(0)}(x, \dot x, \l)\over \d \dot x}\dot x^{(1)}.
\ee
It is easy to obtain the action $S^{(1)}$ for this new Lagrangian and 
write the equations of motion. These turn out to be
$$ 
{\v S^{(1)}\over \v x^{(0)}(t)}=0\qquad\Rightarrow$$
\be  \label{*}
{\d^2L^{(0)}\over \d x^{(0)}\d \l}+ {\d^2L^{(0)}\over [\d x^{(0)}]^2}x^{(1)}
+{\d^2L^{(0)}\over \d x^{(0)} \d \dot x^{(0)}}\dot x^{(1)}
-{{\rm d}\over {\rm d}t}\Big\{{\d^2L^{(0)}\over \d \dot x^{(0)}\d \l}
+{\d^2L^{(0)}\over \d x^{(0)} \d \dot x^{(0)}} x^{(1)}
+{\d^2L^{(0)}\over [\d \dot x^{(0)}]^2}\dot x^{(1)}\Big\}=0, 
\ee
and 
\be \label{**}
{\v S^{(1)}\over \v x^{(1)}(t)}=0  \qquad \Rightarrow\qquad 
{\d L^{(0)}\over \d x^{(0)}}- {{\rm d}\over {\rm d}t}{\d L^{(0)}
\over \d \dot x}=0.
\ee
It is obvious that the equation (\ref{**}) is in fact the Euler-Lagrange 
equation obtained from $S^{(0)}$. One also notices that equation 
(\ref{*}) is the total derivative of (\ref{**}) with respect to $\l$.
That is 
\be \label{*1}
{\v S^{(1)}\over \v x^{(0)}(t)}={{\rm d}\over{\rm d}\l}{\v S^{(0)}\over 
\v x^{(0)}(t)}=0.
\ee

We can repeat this procedure, differentiate $S^{(0)}$, $n$ times and 
obtain $S^{(n)}$:
\be
S^{(n)}:={{\rm d}^n\over{\rm d}\l^n} S^{(0)},
\ee
and then write the Euler-Lagrange equations. It can be easily shown that 
\be
{\v S^{(n)}\over \v x^{(k)}(t)}={n\choose k}
{{\rm d}^{n-k}\over{\rm d}\l^{n-k}} {\v S^{(0)}\over \v x^{(0)}(t)}.
\ee
So, we obtain $n+1$ equations of motion. Note that this differentiation of 
the action does not alter the previous equation of motions. It just adds 
one more equation and one more variable. Also notice that the equation of 
motion of $x^{(0)}$ contains just $x^{(0)}$, and the equation of motion 
$x^{(i)}$
\be
{\v S^{(n)}\over \v x^{(n-i)}(t)}=0
\ee
contains  $x^{(0)}$, $x^{(1)}$, $\cdots$, $x^{(i)}$. That is, these equations 
are recuresive: one obtains $x^{(0)}$ first, then uses this to obtain 
$x^{(1)}$, and so on. 

Now suppose a general solution of the equation (\ref{*}) has been obtained, 
that is,
\be \label{1!}
x^{(0)}(t)=x^{(0)}(t,a,b,\l) 
\ee
satisfies (\ref{*}) for arbitrary values of $a$ and $b$. Treat $a$ and $b$ 
as functions of $\l$, and differentiate (\ref{*1}) with respect to $\l$; 
one obtains:
\be  \label {2*}
x^{(1)}(t)=
{{\rm d}a\over {\rm d} \l}{\d x^{(0)}\over \d a}
+{{\rm d}b\over {\rm d} \l}{\d x^{(0)}\over \d b}
+{\d x^{(0)}\over \d \l},
\ee
which obviously satisfies (\ref{*1}). Note, however, that in 
this function there are two additional arbitrary constants
${{\rm d}a/(\rm d\l )}$ and ${{\rm d}b/(\rm d\l )}$.
So the solutions (\ref{1!}) and (\ref{2*}) have sufficient 
constants of integration  which means that these are in fact the most 
general solution of the equations of motion (\ref{**}) and (\ref{*1}).
One can generalize this procedure up to $x^{(n)}(t)$. In this case, there 
appears, in the solution of the Euler-Lagrange equations, constants 
of integration 
$(a, b, {{\rm d}a/({\rm d}\l )}, {{\rm d}b/({\rm d}\l )}, \cdots
{{\rm d}^na/({\rm d}\l )^n}, {{\rm d}^nb/({\rm d}\l )^n})$,
which are sufficient to provide a general solution. Note that as 
we are considering $S^{(n)}$ at a fixed value of $\l$, the dynamical 
variables $x^{(i)}$, and the constants ${\rm d}^ia/({\rm d}\l )^i$ 
and ${\rm d}^ib/({\rm d}\l )^i$ are independent.

To summerize, we begin with a system of $f$ degrees of freedom, 
and obtain another system with $(n+1)f$ degrees of freedom. The appropriate 
Lagrangian, equation of motion, and their solutions are systematically
derived from the corresponding entities of the initial problem.
\vskip 10mm
\section{\bf Hamiltonian formulation}
Consider the Lagrangian
\be \label{2**}
L_{\D }^{(n)}={1\over \D ^n}\sum_{i=0}^n (-1)^{n-i}{n\choose i} 
L(q^i,\dot q^i, \l+i\D ).
\ee
Defining 
\be \label{*2*}
x_{\D }^{(k)}={1\over \D^k}\sum_{i=0}^k (-1)^{k-i}{k\choose i}q^i,
\ee
it is seen that in the limit $\D\to 0$ (\ref{2**}) and (\ref{*2*}) 
tend to the 
definitions of the previous section, provided that one treats $q^i$ formally
as $q(\l+i\D )$. The momenta conjugate to the coordinates $q$, are defined as 
\bq
\pi_i :&\ \f =&\f {\d L_{\D }^{(n)}\over \d \dot q^i}\nonumber\\
&\ \f =&\f { (-1)^{n-i}\over \D^n }{n\choose i}
{\d L(q^i,\dot q^i,\l+i\D )\over \d \dot q^i}\nonumber\\ 
&\ \f =&\f { (-1)^{n-i}\over \D^n}{n\choose i}\hat \pi_i
\eq
Writing the point transformation (\ref{*2*}) as
\be 
x_{\D }^{(k)}=\L ^k{}_iq^i,
\ee
one can construct a corresponding canonical transformation:
\be
p^{(n)}_{k\D}=\pi_i(\L ^{-1})^i{}_k,
\ee
(as $\L $ is independent of $q$). The inverse matrix is easily seen
to be
\be
(\L ^{-1})^i{}_k=\D^k{i\choose k}.
\ee
So, one obtains
\bq
p_{k\D}^{(n)}&\f =&\f \sum_{i=k}^n{(-1)^{n-i}\over\D^{n-k}}{n\choose i} 
{i\choose k}\hat \pi_i\nonumber\\ 
&\f =&\f \sum_{i=k}^n{(-1)^{n-i}\over\D ^{n-k}}{n\choose k}{n-k\choose i-k}
\hat \pi_i, 
\eq
or
\be
p_{k\D }^{(n)}={n\choose k}\sum_{j=0}^{n-k}{(-1)^j\over \D ^{n-k}} 
{n-k\choose j}\hat \pi_{n-j}.
\ee
It is seen that, as functions of the configuration space variables,
\be
p^{(n)}_k={n\choose k}{{\rm d}^{n-k}\over {\rm d}\l^{n-k}} p
\ee
where the left hand side is the limit of $p_{\D}$ as $\D \to 0$, and
\be \label{**2}
p(x^{(0)},\dot x^{(0)}):={\d L^{(0)}\over \d \dot x^{(0)}}.
\ee
Note that the functional dependence of the conjugate momentum $p^{(n)}_i$ 
of the coordinate $x^{(i)}$, does depend on $n$, the number of 
differentiations, as it is seen from (\ref{**2}).
Using this one can construct the Hamiltonian corresponding to the Lagrangian
$L^{(n)}$. We have
\bq
H^{(n)}&\f =&\f \sum_i p^{(n)}_i \dot x^{(i)} -L^{(n)}\nonumber\\
       &\f =&\f \sum_i {n\choose i}\Big({{\rm d}^{n-i} p\over {\rm d}\l^{n-i}}
       \Big)\Big({{\rm d}^i\dot x\over {\rm d}\l^i}\Big)
       -{{\rm d}^n L\over {\rm d}\l^n }\\
       &\f =&\f {{\rm d}^n \over {\rm d}\l^n } (p\dot x -L)\nonumber
\eq 
or 
\be
H^{(n)}={{\rm d}^n H\over {\rm d}\l^n }
\ee
Once again, note that this relation holds, provided one writes it in 
terms of the variables of the configuration space. 

Now starting from the definition (\ref{2**}), for $L^{(n)}_{\D}$.
We have in fact $n+1$ Lagrangian in terms of $q$'s each depending on just 
one of $q^i$'s. From each Lagrangian, one can construct a Hamiltonian:
\be
h_i:=\dot q^i \hat \pi_i -L(q^i, \hat q^i, \l +i\D)
\qquad {\rm no \ \ summation}
\ee
Using these, we define 
\be \label{3**}
H_{k\D }^{(n)}={1\over \D^k}\sum_{i=0}^k (-1)^{k-i}{k\choose i}h_i
\ee
In the limit $\D\to 0$, one obtains 
\be \label{*3*}
H^{(n)}_k={{\rm d}^k H\over {\rm d}\l^k }
\ee
Unlike the momenta, the functional dependence of these on the 
variables of the configuration space does not depend on $n$. However, their 
dependence on the variables of the phase space, does depend on $n$. 
Now writng $h_i$'s in terms of $q$ and $\pi$, it is seen that 
\be 
\{ h_i,h_j\}_{q,\pi}=0
\ee
This relation, among with the definition ({\ref{3**}), leads to 
\be 
\{ H_{i\D}^{(n)}, H_{j\D}^{(n)}\}_{q,\pi}=0
\ee
However, as a canonical transformation does not change the poisson bracket  
structure, the above relation is also true in terms of $x$ and $p$.
Then, letting $\D\to 0$, we find that  
\be 
\{ H^{(n)}_i,H^{(n)}_j\}_{x,p}=0,
\ee
The importance of this relation is that one authomatically obtains $n+1$
involutive constants of motion.

At last we come to the question of obtainig $H^{(n)}$ and $H^{(n)}_i$'s
directly from the form of $H$ in the phase space. Regarding (\ref{**2})
and (\ref{*3*}) it is easy to see that first, one must change 
\be
H(x,p)\to H(x^{(0)},p^{(n)}_n)=H^{(n)}
\ee
Then differentiate it $k$ times to obtain 
\be
H^{(n)}_k={{\rm d}^k \over {\rm d}\l^k } H(x^{(0)},p^{(n)}_n)
\ee
and use (\ref{**2}) to obtain expressions for the derivative of $p^{(n)}_n$.
In this way we obtain $n+1$ functions of the phase space, the last of which is 
the Hamiltonian of the system:
\be
H^{(n)}_n=H^{(n)}= {{\rm d}^n \over {\rm d}\l^n }  H(x^{(0)},p^{(n)}_n).
\ee
\vskip 10mm
\section{\bf Symmetries and constants of motion}
The formulation of the previous section can be easily used to treat 
symmetries. A symmetry, in general is an operation which changes an 
arbitrary solution of the equation of motion to another solution. 
Suppose that the system defined through $L$, relation (1),
is symmetric under the action of ${\cal O}$:
\be
{\cal O} :x(t)\to ({\cal O} x)(t)
\ee
Now, consider the system defined by $L^{(1)}$. The dynamical variables 
of this system are $x^{(0)}$ and $x^{(1)}$. Using the discretization 
procedure of the previous section, and the variables $q^0$ and $q^1$,
one can see that any solution $(q^0(t),q^1(t))$ of the equations of motion
change to another solution $({\cal O }q^0, {\cal O }q^1)$, under the action 
of ${\cal O}$. However,
\be 
q^1=q^0+\D x_{\D}^{(1)}
\ee
So one can define 
\be
{\cal O}x_{\D}^{(1)}={1\over \D}[{\cal O}(q^0+\D x_{\D}^{(1)})-{\cal O}(q^0)]
\ee
or
\be 
{\cal O} x^{(1)}={{\rm d}\over {\rm d}\l}[{\cal O} x^{(0)}]
\ee
Proceeding in this way, one can define the action of ${\cal O}$ on 
the dynamical variables through
\be \label{***}
{\cal O} x^{(i)}={{\rm d}^i\over {\rm d}\l^i}[{\cal O} x^{(0)}].
\ee
This action has the property that changes any solution of the equations 
of motion to another solution, so it is a symmetry of the new system. 

The symmetry ${\cal O}$ is called Notherian if it does not change the 
Lagrangian, or changes it by a total derivative. It is straight 
forward to see that if ${\cal O}$ is a Notherian symmetry of $L$, then
it is also the Notherian symmetry of the $L^{(n)}_{\D}$.
Letting $\D \to 0$, we calculate that ${\cal O}$, the definition of its 
action extended through  (\ref{***}), is also a Notherian symmetry of 
$L^{(n)}$. Any Notherian symmetry which is continiously related to identity
results a conserved quantity (according to Nother's theorem).
Now we can show that any infinitesimal Notherian symmetry of $L$, results 
in $n+1$ conserved quantities for the system described by the Lagrangian
$L^{(n)}$. In fact, if the symmetry action 
\be
x\to x+ \e {\cal G}
\ee
where ${\cal G}$ is the generator of the symmetry which may depend on 
$ x, \l$ and $t$, and under the action of this operator, 
\be
L\to L+\e {{\rm d}f\over {\rm d}t},
\ee
then it is easy to check that under 
\be
x^{(i)}\to x^{(i)}+ \e {{\rm d}^i{\cal G}\over {\rm d} \l^i},
\ee
the Lagrangian $L^{(n)}$ is transformed according to 
\be
L^{(n)}\to L^{(n)}+ \e {{\rm d}^n\over {\rm d} \l^n}
\Big({{\rm d}f\over {\rm d}t}\Big) ,
\ee
So that this is a Notherian symmetry. The conserved 
quantity corresponding to this symmetry is 
\bq
I^{(n)}
       &\f =&\f \sum_i p^{(n)}_i {{\rm d}^i{\cal G}\over {\rm d} \l^i}
        -{{\rm d}^nf\over {\rm d} \l^n}\nonumber\\
       &\f =&\f \sum_i{n\choose i}{{\rm d}^{n-i}p\over {\rm d} \l^{n-i}}
        {{\rm d}^i{\cal G}\over {\rm d} \l^i}
        -{{\rm d}^nf\over {\rm d} \l^n}
\eq
or 
\bq
I^{(n)}&\f =&\f {{\rm d}^n\over {\rm d} \l^n} (p{\cal G}-f)\nonumber\\
       &\f =&\f {{\rm d}^n\over {\rm d} \l^n} I
\eq
However, note that not only $I^{(n)}$, but also ( for $0\leq i\leq n$ )
\be
I^{(n)}_i:={{\rm d}^i\over {\rm d} \l^i} I
\ee
is constant as well. The proof of this is similar to the proof of the 
constancy of the Hamiltonian. In other words, any Notherian symmetry 
of the initial system results in a Notherian symmetry of the final system
and $n+1$ constants of motion. There may be  a constant of motion not coming 
from the Notherian symmetry but from a dynamical symmetry. It is easily seen 
that if $C(x,\dot x , \l ,t)$ is such a constant of motion of the initial 
system, then 
\be
C^{(n)}_i={{{\rm d}^i}\over{{\rm d}\lambda^i}}
\ee
is a constant of motion of the final system. So beginning by a constant of 
motion of initial system, we end by $n+1$ constants of motion of the final
system. The functional dependence of $C^{(n)}_i$'s on the variables of the 
configuration space does not depend on $n$, and $C^{(n)}_i$ depends on the 
configuration space variables up to $x^{(i)}$ and $\dot x^{(i)}$. 
But in phase space, this functional dependence does depend on $n$, as it was 
the case for the Hamiltonian.
\vskip 10mm
\section{\bf Integrability}

A system of $f$ degrees of freedom is integrable if it has $f$ involutive 
( and independent ) constants of motion $\pi_{\a}(\a =1,2,\cdots,f)$:
\be
\{\pi_{\a}, H\}=\{\pi_{\a},\pi_{\b}\}=0.
\ee
This is the phase space description. For the configuration space 
description,  one can can say that the system is integrable if one 
can find well-behaved function(s) of time and initial values, (\ref{1*}),
which satisfy the equation(s) of motion. For the configuration space,
we saw that integrability of $L$ results in the integrability of $L^{(n)}$.
In phase space, we want to obtain also $(n+1)f$ involutive constants of 
motion. We have two choices. The first one is to express $\pi$'s in terms  
of $q$'s and $\dot q$'s. Then use the dicretization procedure, obtain 
$\pi_{\D i\a}$'s and at last write derivatives of $\pi_{\a}$'s in terms of 
$x$'s and $\dot x$'s. At the first stage, one can express $\dot x$'s in
terms of $p$'s. This is similar to the procedure done for the Hamiltonian.
The same reason also works here to show that these quantities are in fact 
involutive.

The more straight forward procedure is to use (21)  and change 
\be
\pi_{\a}(x,p)\to \pi_{\a}(x^{(0)},p^{(n)}_n)=P_{\a}^{(n)}(x^{(0)},p^{(n)}_n)
\ee
Now, differentiate it with respect to $\l$ to obtain 
\be
P_{\a i}^{(n)}={{\rm d}^i\over {\rm d}\l^i}P_{\a 0}^{(n)}(x^{(0)},p^{(n)}_n)
\ee
In this way , one obtains $(n+1)f$ involutive independent constants of motion which show the 
integrability of the system described by $H^{(n)}$
\vskip 10mm
\section{A simple example}
Consider the simple Lagrangian of a harmonic oscilator,
\be
L={m\over 2}\dot x^2-{m \o^2\over 2}x^2=:L^{(0)}
\ee
Treat $\o$ as the variable parameter and obtain 
\be
L^{(1)}= m\dot x^{(0)}\dot x^{(1)} -m\o [x^{(0)}]^2 -m\o ^2x^{(0)}x^{(1)}.
\ee
The equation of motion corresponding to $L^{(1)}$ are
\be \label{****}
{\v S^{(1)}\over \v x^{(1)}}=-m\ddot x^{(0)}  -m\o ^2 x^{(0)} =
{\v S^{(0)}\over \v x^{(0)}}=0
\ee
and 
\be \label{**4}
{\v S^{(1)}\over \v x^{(0)}}=-m\ddot x^{(1)}  -m\o ^2 x^{(1)}  -
2m\o x^{(0)}=0
\ee
which is the derivative of the first equation.
According to the discretization procedure of section 3, we could suspect 
that this system in fact consists of two independent harmonic oscilators.
This is, however, true only when we have not let $\D \to 0$.
The system we are considering has another feature. To see this solve 
(\ref{****}) to obtain 
\be \label{4*}
x^{(0)}=a \cos \o t +b\sin \o t
\ee
It is seen that 
\be
x^{(1)}=a' \cos \o t +b'\sin \o t -ta \sin \o t +tb\cos\o t
\ee
which could be obtained by differentiation (\ref{4*}) with respect to $\o $.

One can suggest a physical system where its equations of motion are like 
(\ref{****}) and (\ref{**4}). Suppose that we have two masses $m_1$ and
$m_2$, and three springs $k_1,k_2$ and $k_3$. $k_1$ is attached to a wall 
and $m_1$, $k_2$ to $m_1$ and $m_2$, and $k_3$ to $m_2$ and another 
wall. The whole system is contained in one dimension.
Denoting the positions of $m_1$ and $m_2$ by $x$ and $y$, respectively,
we arrive at the equations of motion 
\bq \label{4**}
\ddot x +{k_1+k_2\over m_1}x-{k_2\over m_1}y&\f =&\f 0\nonumber\\
\ddot y +{k_2+k_3\over m_2}y-{k_3\over m_2}x&\f =&\f 0
\eq 
Now suppose that 
\be 
{k_1\over m_1}={{k_2+k_3}\over m_2}=\o^2 \qquad k_2<<k_1 \quad (m_2<<m_1 )
\ee
the equations  (\ref{4**})then are simplified:
\bq 
\ddot x+\o^2 x&\f =&\f 0\nonumber\\
\ddot y+\o^2 y-{k_3\over m_2}x&\f =&\f 0
\eq
But these are just (\ref{****}) and (\ref{**4}), provided one defines
\be 
x^{(0)}=x \qquad x^{(1)}=-{2m_2\o \over k_3}y
\ee
This system contains of a very massive oscilator, which oscilates with the 
natural frequencyof another oscilator, thus putting this second oscilator 
in resonance.

In phase space we have 
\be
H={P^2\over 2m} ={m\o^2 \over 2} x^2
\ee
which gives 
\be 
H^{(1)}_0={[P^{(1)}_1]^2\over 2m} +{m\o^2 \over 2}[x^{(0)}]^2
\ee
differentiating this, we obtain 
\be 
H^{(1)}_1={P^{(1)}_1P^{(1)}_0\over m} +m\o^2 [x^{(0)}]^2 +m\o^2 x^{(0)} 
x^{(1)}
\ee
It is easy to check that this two functions ( the second of which is the
Hamiltonian ) are in fact involutive. So that the system is integrable
( as it should be ).

\section{Field theory and the Lax pair}

It is obvious that the above constructions is applicable also to any 
classical field theory. Also note that if the original field theory is 
integrable, the derived one is also integrable. Now, if a $1+1$ dimensional 
field theory has a Lax structure, that is, if there exists a connection 
in terms of the field so that the zero curvature condition of this 
connection is equivalent to the equation of motion of the field, then the 
theory is integrable. The zero curvature condition is written as
\be\label{99} [D_0,D_1]=0,\ee
where
\be D_\mu :=\d _\mu +L_\mu\qquad\mu=0,1.\ee

Suppose that the original theory has a Lax structure. The equation 
(\ref{99}) is then written as
\be\label{98}\d_0 L_1-\d_1 L_0+[L_0, L_1]=0.\ee
Differentiating this with respect to $\lambda$, we obtain
\be\label{97}\d_0 L'_1-\d_1 L'_0 +[L'_0, L_1] + [L_0, L'_1]=0,\ee
where
\be L'_\mu :={{\rm d}\over{{\rm d}\l}}L_\mu .\ee
This equation is of course equivalent to the equation of motion of the 
derived field. It is not difficult to see that equations (\ref{98}) and 
(\ref{97}) can be combined in the single equation 
\be\d_0 L^{(1)}_1-\d_1 L^{(1)}_0+[L^{(1)}_0, L^{(1)}_1]=0,\ee
where
\be\label{96}L^{(1)}_\mu :=\pmatrix{L_\mu&0\cr L'_\mu&L_\mu\cr}.\ee
This means that if the original theory has a Lax structure, the derived 
theory has a Lax structure as well. The Lax pair of the derived theory is 
expressed in terms of that of the original theory through (\ref{96}).

\noindent{\bf Acknowledgement} We would like to thank M. Alimohammadi, 
M. R. Rahimi Tabar, and A. Shariati, for useful discussions.

\newpage

\end{document}